# T2 estimation from MOLLI acquisitions

**Catarina N. Carvalho[1, 2]*, Andreia S. Gaspar [2], Rita G. Nunes [2], Teresa M. Correia [1, 3]**

[1] Quantitative Bio-Imaging Lab, Centre of Marine Sciences (CCMAR), Universidade do Algarve, Campus de Gambelas, Faro, Portugal
[2] Institute for Systems and Robotics - Lisboa and Department of Bioengineering, Instituto Superior Técnico – Universidade de Lisboa, Lisbon, Portugal
[3] School of Biomedical Engineering and Imaging Sciences, King's College London, London, United Kingdom

***Corresponding author**: tmcorreia@ualg.pt

**September 2026**

## Abstract

Cardiac $T_1$ and $T_2$ quantitative magnetic resonance imaging (MRI) is a technique that provides characterization of a multitude of myocardial pathologies. However, each parameter requires its own specialized acquisition sequence, which may be time-consuming. In this work, we demonstrate that MOLLI, an acquisition sequence commonly used to perform $T_1$ mapping, is also sensitive to $T_2$, and can also be used for myocardial $T_2$ mapping. Although MOLLI is not designed for $T_2$ mapping, its signal evolution inherently depends on both $T_1$ and $T_2$ relaxation. Here, we demonstrate that this dependence is sufficiently strong to allow $T_2$ estimation directly from MOLLI $T_1$-weighted images, using appropriate signal modelling. This finding opens the possibility of obtaining simultanesouly $T_1$ and $T_2$ maps from a single MOLLI acquisition, reducing total scan time and potentially improving patient comfort and clinical throughput.

## Keywords

Quantitative Magnetic Resonance Imaging, Relaxometry, T1 mapping, T2 mapping, MOLLI acquisition

## 1. Introduction

Quantitative Magnetic Resonance Imaging (MRI) is a technique that provides quantitative assessment of inherent tissue properties whilst taking advantage of increased sensitivity to different pathologies, less dependence on hardware variations, acquisition settings and operator expertise, and improved objectivity and reproducibility [1–3]. The most frequently mapped parameters in cardiac MRI are $T_1$, $T_2$, $T_2^*$ and extracellular volume [4].

In particular, the native $T_1$ and the $T_2$ of the myocardium are important biomarkers in clinical practice that allow clinicians to characterize the myocardium's local molecular environment and identify abnormal areas, providing characterization of multiple pathologies such as myocardial infarction, myocarditis, cardiac amyloidosis, iron overload, edema and myocardial hemorrage [5, 6].

Myocardial $T_1$ mapping is usually performed with a Modified Look-Locker Inversion Recovery (MOLLI) acquisition sequence, due to its high Signal to Noise Ratio (SNR), good reproducibility and $B_1$ insensitivity [5]. Meanwhile, myocardial $T_2$ mapping is accomplished with other distinct sequences, such as $T_2$-prepared Balanced Steady State Free Precession (bSSFP) acquisitions [7].

Despite its advantages, quantitative mapping sequences are time-consuming and sensitive to motion, since they require the acquisition of multiple weighted-images for posterior signal modelling and parameter estimation. In this work, we demonstrate that MOLLI is also sensitive to $T_2$ and can be used to simultaneously obtain myocardial $T_2$ map estimates from a single $T_1$ acquisition sequence, potentially speeding up cardiac MRI examinations.

# 2. Methods

## 2.1 Data Acquisition

The $T_1$ and $T_2$ mapping CMRxRecon dataset was used [8]. Subjects #001, #004, #007, #013 and #028 were included in the analysis, corresponding to Heart Rates (HRs) of 75, 85, 55, 65 and 94 bpm, respectively.

**$T_1$ mapping sequence.** The acquisition protocol consisted of a MOLLI sequence (4-(1)-3-(1)-2) acquired in Short Axis (SA) view only, Field of View (FOV)=360×307 $mm^2$, spatial resolution=1.4×1.4 $mm^2$, slice number 5 ~6, slice thickness=5.0 mm, TR=2.67 ms, TE=1.13 ms, 32 coils, partial Fourier=7/8, and GeneRalized Autocalibrating Partial Parallel Acquisition (GRAPPA) factor =2, on a 3T Vida Siemens scanner.

**$T_2$ mapping sequence.** The acquisition protocol consisted of a $T_2$-prepared Fast Low Angle Shot (FLASH) sequence with three $T_2$ weightings in SA view, with identical geometrical parameters as used in $T_1$ mapping, TE = 1.29 ms, and $T_2$ preparation time = 0/35/55 ms.

Coil compression was used (10 virtual coils) and data were reconstructed prior to the analysis.

## 2.2 Signal Modelling

$T_1$ and $T_2$ modelling followed the Extended Phase Graph (EPG) formulation [9]. EPG dictionaries consisting of the simulated signal evolution curves were generated, for all combinations of $T_1 \in \{500, 501, 502, \ldots, 2498, 2499, 2500\}$ ms and $T_2 \in \{30, 31, 32, \ldots, 348, 349, 350\}$ ms, following the inversion pulse timings of the 5 distinct subjects referenced above, with HRs equal to 55, 65, 75, 85, and 94 bpm.

$T_1$ and $T_2$ mapping for each subject were performed through dictionary matching, i.e., choosing the combination of $T_1$ and $T_2$ that maximizes the dot product between the corresponding signal evolution curve and the measured signal, for each pixel. Additionally, $T_2$ mapping was also conducted through mono-exponential fitting, following the signal model

$$S(\mathbf{x}, t) = \mathrm{PD}(\mathbf{x})\exp\left(-\frac{t}{T_2(\mathbf{x})}\right), \tag{1}$$

where $S(\mathbf{x}, t)$ represents the magnitude of the signal for the spatial position $\mathbf{x}$ at time $t$, PD and represents the proton density.

## 2.3 Simulated Dataset

In order to evaluate MOLLI's sensitivity to $T_2$ with a known ground-truth, a simulated dataset was generated.

### 2.3.1 Generating ground-truth quantitative maps

An example phantom from MRXCAT [10] was selected for simulations, where each tissue is masked independently. To generate a set of $T_1$ and $T_2$ parametric maps, for each pixel a $T_1$ and $T_2$ value were sampled from a normal distribution with means $\mu_{T_1}$=$\{1800, 1100\}$ ms, $\mu_{T_2}$=$\{200, 100\}$ ms and standard deviations $\sigma_{T_1}$=$\{200, 100\}$ ms, $\sigma_{T_2}$=$\{10, 5\}$ ms, in the blood and myocardial regions, respectively. To simulate biological intra-tissue variation in the parametric maps, an additional value was sampled from a normal distribution centred in zero with standard deviation equal to 5% $\mu_{T_1}$, 5% $\mu_{T_2}$:

$$T_1^{\text{GD}}(\mathbf{x}) \sim \mathcal{N}\left(\mu_{T_1}(\mathbf{x}), \sigma^2_{T_1}(\mathbf{x})\right) + \mathcal{N}\left(0, (0.05\mu_{T_1}(\mathbf{x}))^2\right), \tag{2}$$

$$T_2^{\text{GD}}(\mathbf{x}) \sim \mathcal{N}\left(\mu_{T_2}(\mathbf{x}), \sigma^2_{T_2}(\mathbf{x})\right) + \mathcal{N}\left(0, (0.05\mu_{T_2}(\mathbf{x}))^2\right). \tag{3}$$

This procedure was repeated 10 times, resulting in 10 simulated ground-truth parametric map pairs. The final distribution of $T_1$ and $T_2$ values parametric maps across these pairs is illustrated in *figure 1*.

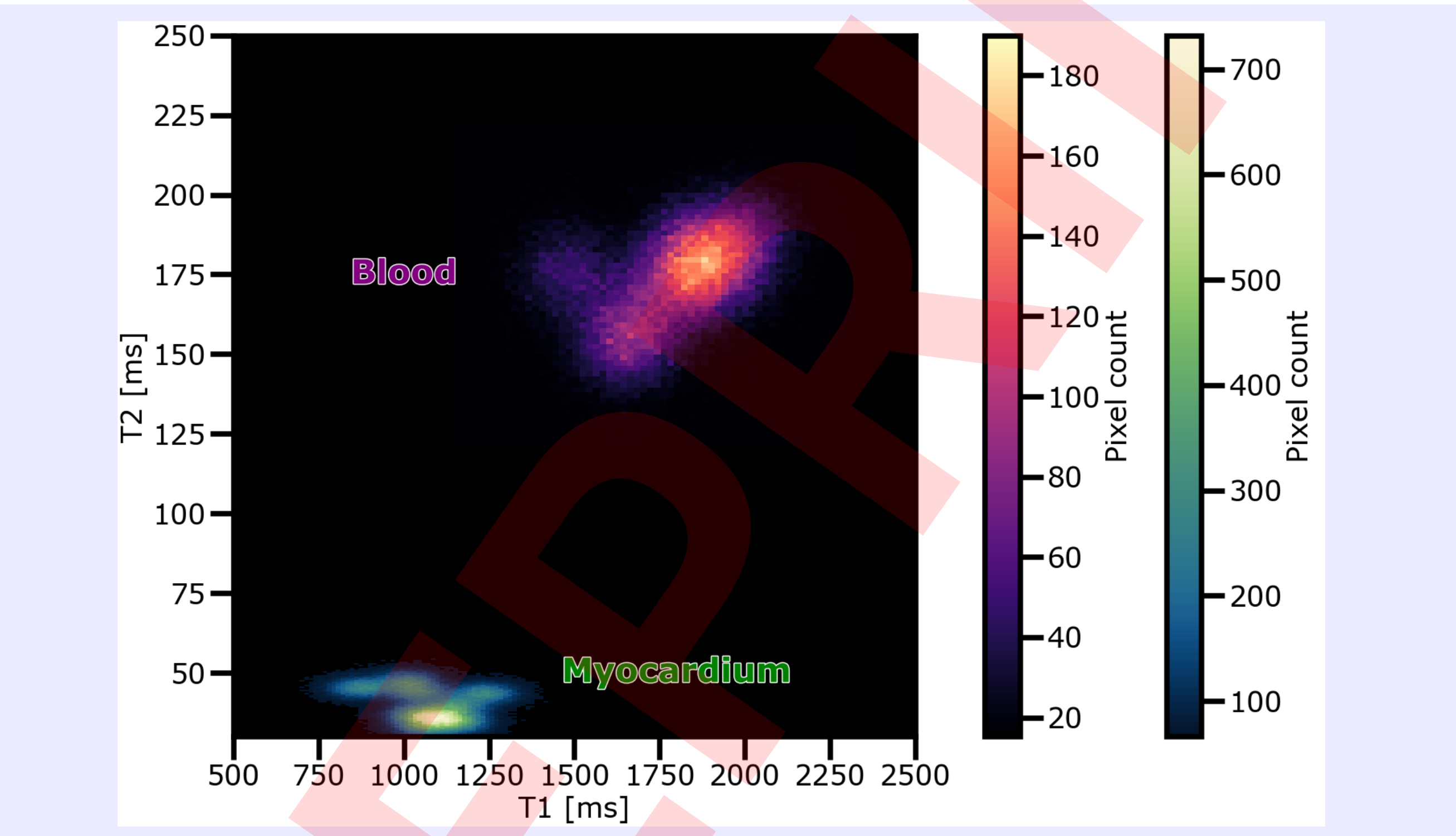

**Figure 1.** 2D histogram of the $T_1$ and $T_2$ value combinations generated for the myocardium and the blood regions, across the 10 simulated parametric map pairs.

### 2.3.2 Generating $T_1$-weighted images

The corresponding signal evolution curves were simulated from each $T_1$, $T_2$ map pair. For each map pair, the corresponding weighted images were extracted from the EPG dictionaries corresponding to each of the 5 subjects selected from the CMRxRecon dataset, resulting in 5 subjects × 10 map pairs = 50 simulated datasets. Additionally, simulated noise was added to the signals following the equation

$$S_N(\mathbf{x}) = S_R(\mathbf{x}) + N_R(\mathbf{x}) + i\left[S_I(\mathbf{x}) + N_I(\mathbf{x})\right], \tag{4}$$

where $S_N$ is the noisy Magnetic Resonance (MR) signal in image space, $i = \sqrt{-1}$, $\mathbf{x}$ is the spatial position where the signal is acquired, $S_R$ and $S_I$ are the real and the imaginary components of the MR signal in the absence of noise, respectively, and $N_R$ and $N_I$ are the real and the imaginary components of the simulated noise [11]. $N_R$ and $N_I$ were sampled from a Gaussian distribution with zero mean and standard deviation $\sigma_N$, where different values of $\sigma_N$ were chosen corresponding to SNRs of 20, 40, 60 and 80. [1] This procedure thus resulted in 50 simulated image sets for each SNR value.

## 2.4 Experiments and Analysis

### 2.4.1 Simulations

The $T_1$ and $T_2$ maps of the simulated dataset estimated with the MOLLI acquisition sequence were visually compared to the corresponding ground-truth maps. A Bland-Altman analysis was then conducted for $T_2$ values in the myocardium and in the blood regions, for SNR $\in \{20, 40, 60, 80, \infty\}$, where $\infty$ corresponds to the noiseless datasets.

To isolate MOLLI's intrinsic sensitivity to $T_2$, we additionally generated controlled 2×2-pixel synthetic image patches with known $T_1$ and $T_2$ values, simulating all combinations of $T_1 \in \{500, 501, ..., 2499, 2500\}$ ms, $T_2 \in \{30, 31, ..., 349, 350\}$ ms, and SNR $\in \{20, 40, 60, 80, \infty\}$, using the acquisition timings of subject #001. The $T_2$ errors of the corresponding estimates with MOLLI were then evaluated for all the image patches.

### 2.4.2 In vivo

The $T_2$ maps estimated in the in vivo dataset with the MOLLI acquisition sequence were visually compared to those estimated using the $T_2$-prepared FLASH sequence. A paired samples t-test was conducted to identify statistically significant differences between methods. Normality was assessed with the Anderson-Darling test.

ROIs were defined for each in vivo dataset, and a Bland Altman analysis was performed to evaluate the relation between the $T_2$ maps estimated from the MOLLI sequence and from the FLASH sequence. The $T_2$ values estimated in the 16-segment American Heart Association (AHA) model with MOLLI and FLASH were also compared to assess whether the methods differ in specific myocardial regions. Significant differences were evaluated with a paired-samples t-test conducted in each region.

# 3. Results and Discussion

## 3.1 Simulations

Illustrative examples of the estimated maps of the simulated datasets are shown in *Figure 2*. As expected, MOLLI's $T_2$ estimation loses some accuracy as the level of noise in the dataset increases. This effect is more pronounced in the blood regions, where the $T_2$ values are larger. However, errors remain lower than 3 ms and 10 ms even for SNR=20, in the myocardium and blood regions, respectively.

A Bland-Altman analysis was conducted in the myocardial and left ventricle blood regions for this dataset (*Figure 3*), demonstrating MOLLI's ability to provide accurate $T_2$ maps, especially in low-noise conditions. While the $T_2$ estimate worsens as the amount of noise in the weighted images increases, MOLLI still provides accurate mean $T_2$ values. In booth the myocardium and the blood, the estimates are comparable with the respective ground-truth values in terms of mean difference, across all noise

[1] SNR is defined as

$$\text{SNR} = \frac{|\overline{S}|}{\sqrt{\frac{2}{4-\pi}}\sigma_N}, \tag{5}$$

where $\overline{S}$ is the mean of the reconstructed MR image pixel intensities in the absence of noise, from a Region of Interest (ROI) in the left ventricle blood pool, and the factor $\sqrt{\frac{2}{4-\pi}}$ accounts for the Rayleigh distribution.

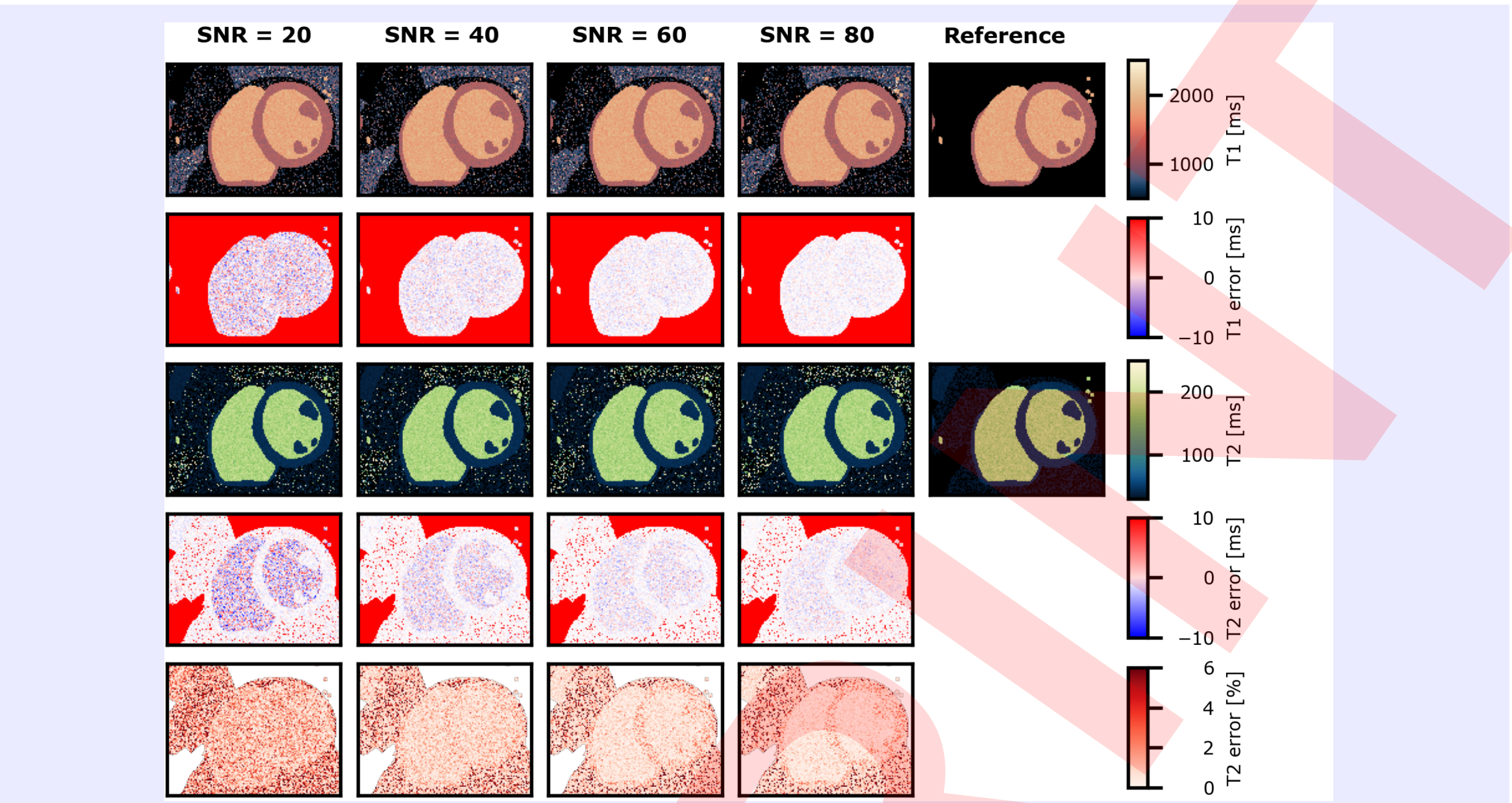

**Figure 2.** Estimated $T_1$ and $T_2$ maps and corresponding error maps from map #1, subject 1 (HR=75), for varying SNR levels.

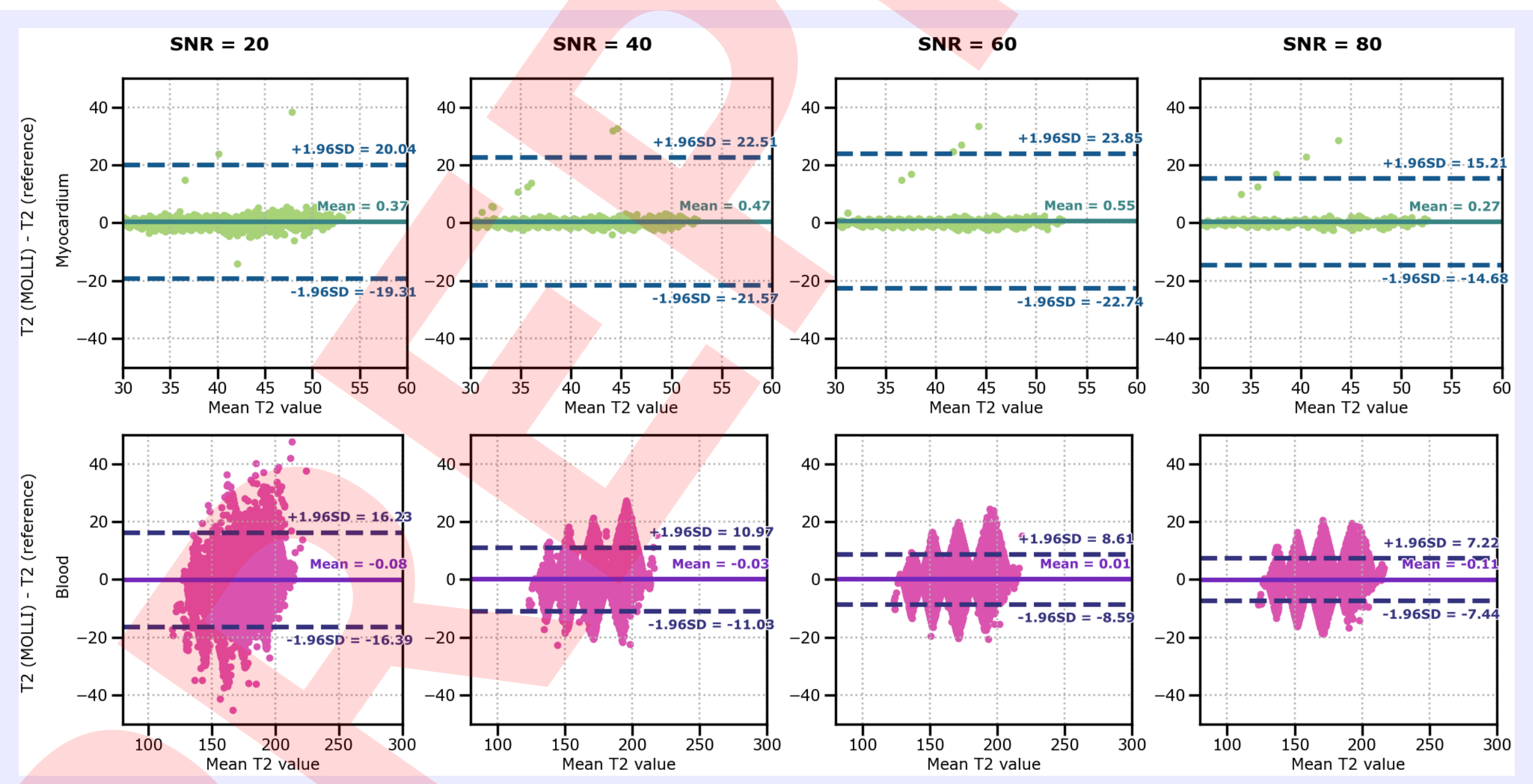

**Figure 3.** Bland-Altman analysis for $T_2$ values estimated with MOLLI compared with the ground-truth, in the myocardium and in the blood regions, for SNR=$\{20, 40, 60, 80\}$.

intensities. As the noise increases, the error distribution widens lightly, an effect that is more significant in the blood region.

We then assessed MOLLI's sensitivity to $T_2$ as a function of $T_2$, $T_1$ and SNR (*Figure 4*). The relative $T_2$ error of the estimates increases with the ground-truth $T_2$ value and decreases with the SNR. This suggests MOLLI may be able to provide accurate $T_2$ myocardial maps in low-noise conditions, where the $T_2$ value is smaller, but not in the blood pool, where the relative errors are expected to be larger. There are no significant variations with $T_1$ value changes.

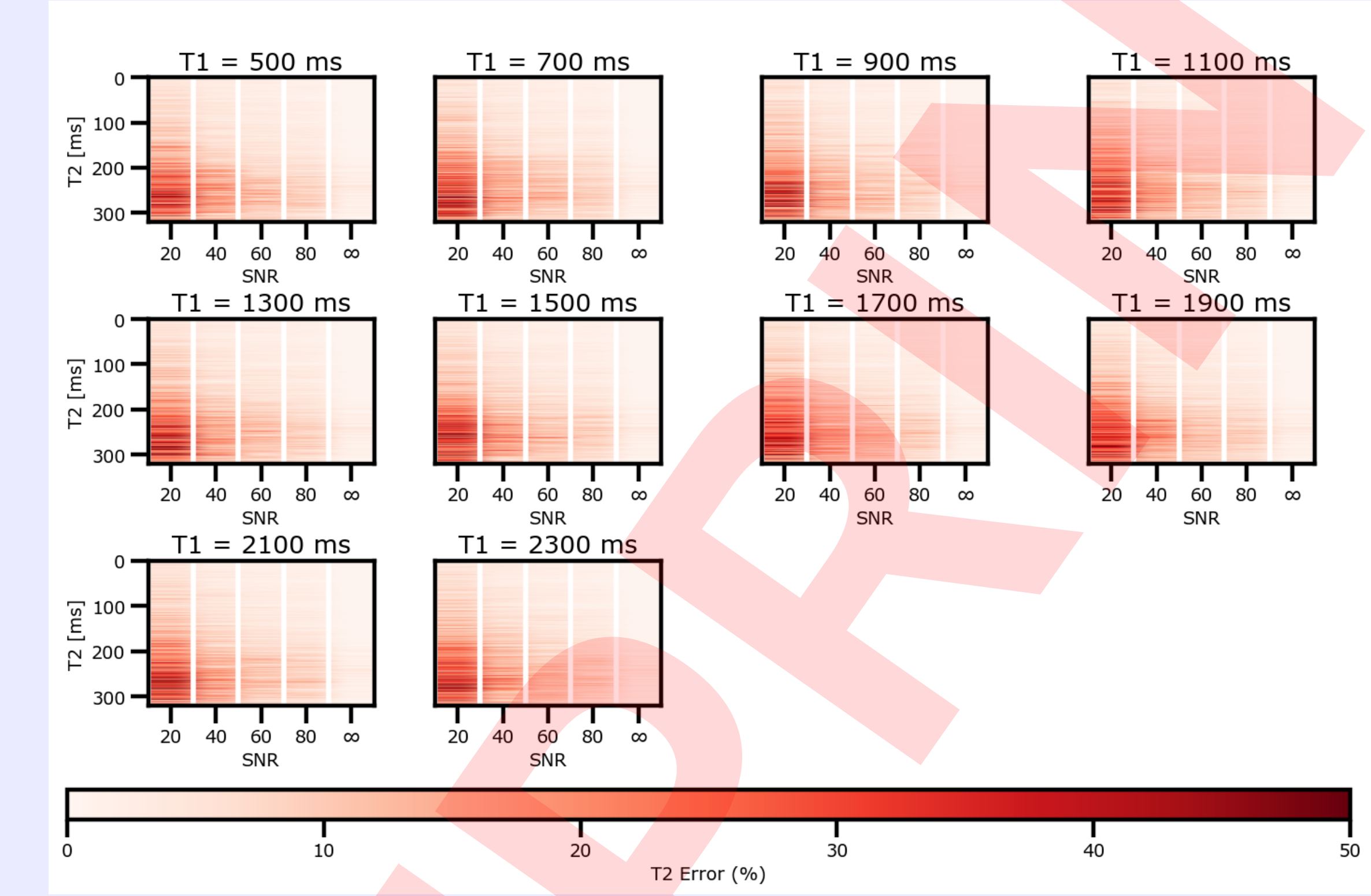

**Figure 4.** Mean relative errors of the $T_2$ maps estimated for $T_2 = \{30, 31, ..., 299, 350\}$ ms, for distinct values of $T_1$, and SNR=$\{20, 40, 60, 80, \text{inf}\}$.

## 3.2 In vivo

The estimated maps for the in vivo datasets are shown in *Figure 5*. There are significant differences between the $T_2$ values estimated from FLASH and MOLLI in the myocardium, and especially in the blood regions (p-value ~ 0), with MOLLI estimating larger $T_2$ values.

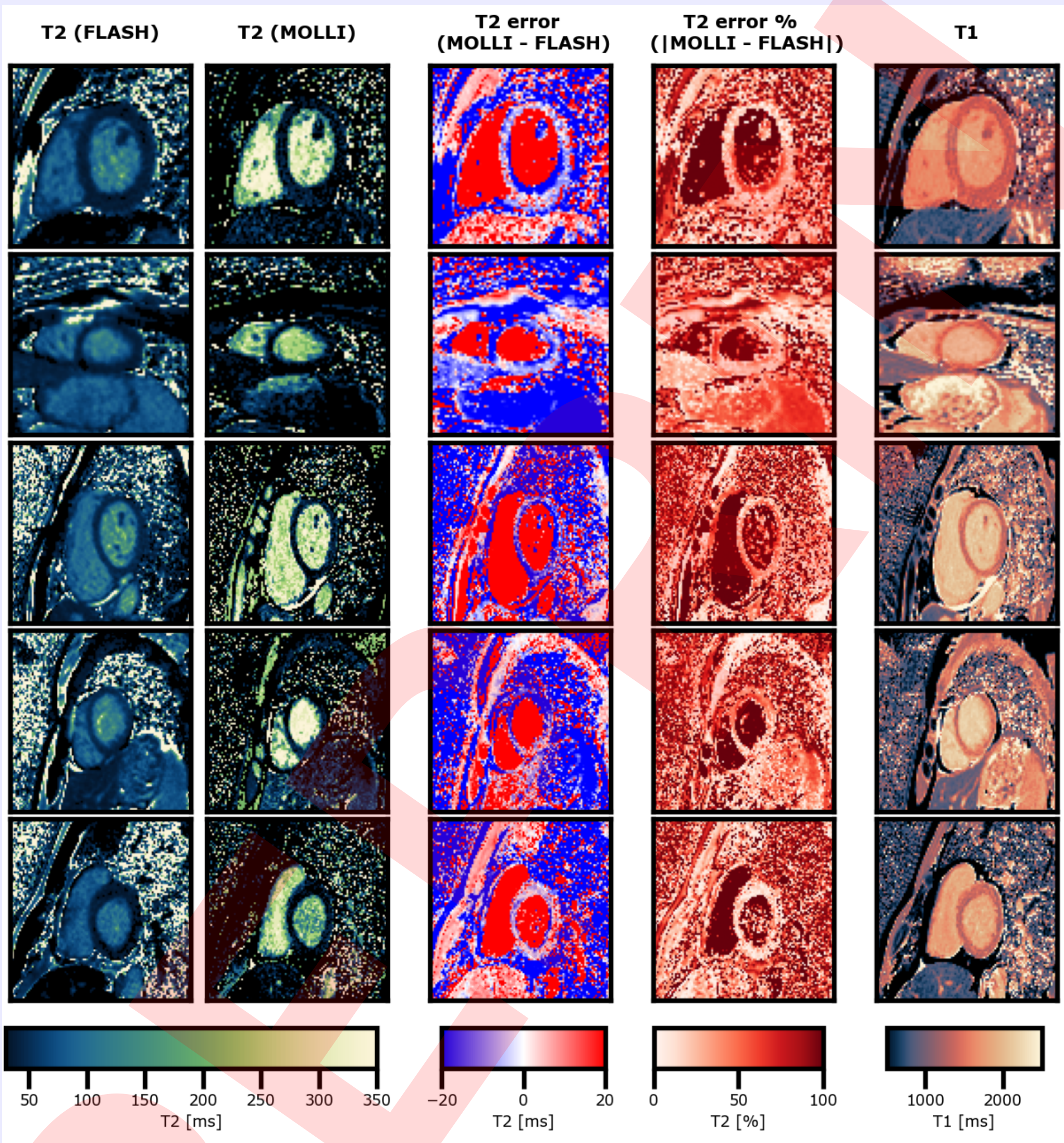

**Figure 5.** Estimated $T_2$ maps from the FLASH sequence and the MOLLI sequence, difference between them, and corresponding estimated $T_1$ maps, for subjects #001 (basal slice), #004 (apical slice), #007 (basal slice), #013 (apical slice) and #028 (basal slice).

A Bland-Altman analysis was conducted (*Figure 6*). In the myocardium , the $T_2$ estimates are comparable in terms of mean difference (-6.26 ms), where the difference may be explained by the different modelling strategies (exponential vs EPG). The distribution of the estimated $T_2$ in each pixel also follows the previous observation that the estimates worsen as the $T_2$ value increases. In the left ventricle blood pool, the estimates of both methods are not comparable, with a mean difference of 144.12 ms, since FLASH $T_2$ mapping is known to be reliable for the myocardium ($T_2 \sim 40$ - 60 ms), but not for the long $T_2$ values characteristic of the blood region.

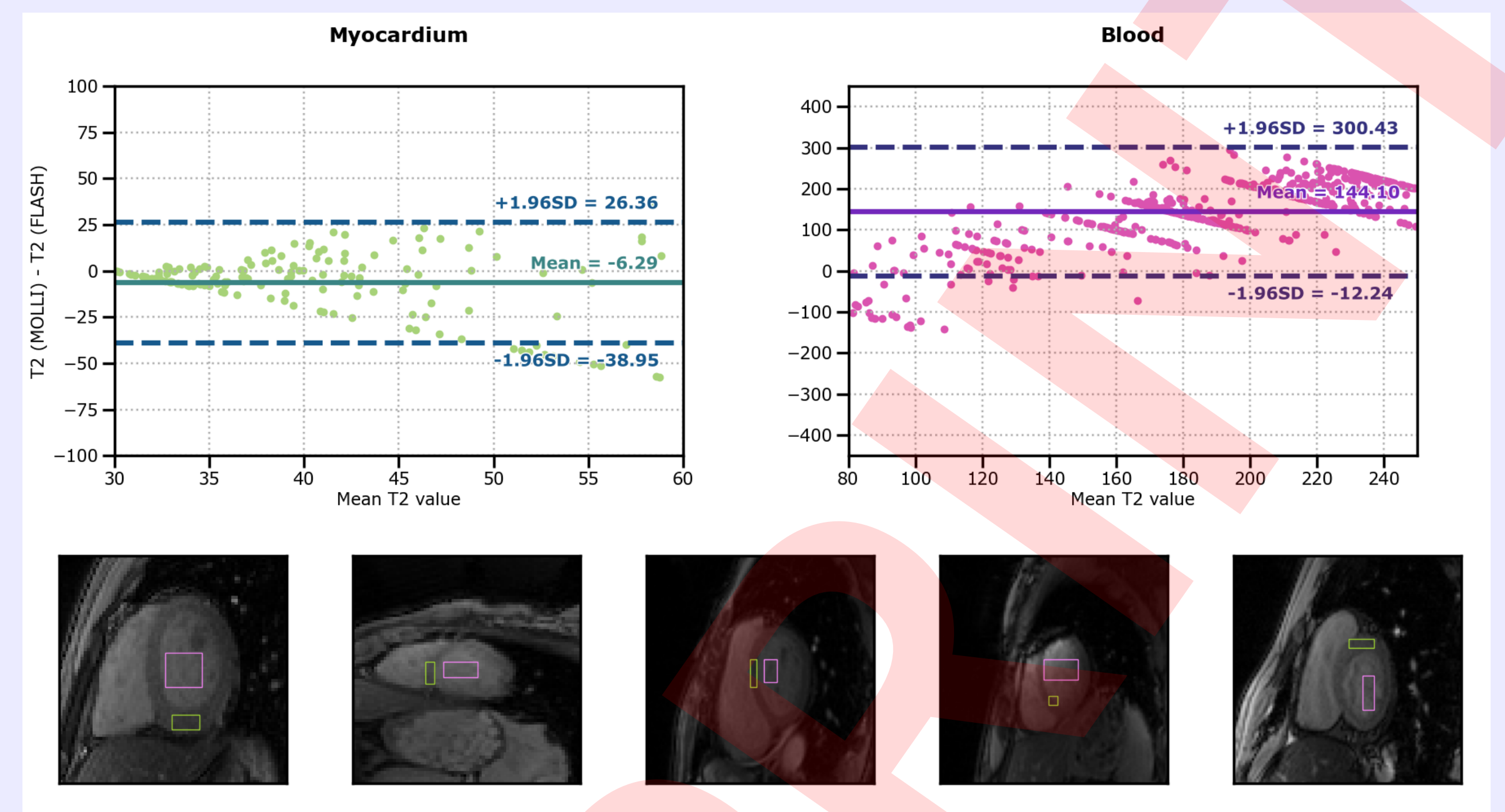

**Figure 6. (top)** Bland-Altman analysis for $T_2$ values comparing FLASH and MOLLI estimates, in the myocardium and in the blood ROIs, for subjects #001 (basal slice), #004 (apical slice), #007 (basal slice), #013 (apical slice) and #028 (basal slice). **(bottom)** Reconstructed first dynamic of each in vivo dataset, where the ROIs are represented.

Finally, to assess whether FLASH and MOLLI $T_2$ estimates differ in specific myocardial regions, the mean $T_2$ values estimated in the 16 AHA segments for both methods as well as the mean standard deviations across all subjects were calculated (*Figure 7*). MOLLI provides $T_2$ myocardial estimates within the expected range. The estimated $T_2$ values for both methods showed the most agreement in the basal septal and inferolateral segments, with errors lower than 2 ms. The apical lateral region showed the largest difference between them (-13.0 ms), with the anterior regions having larger errors overall than other regions. Significant differences between FLASH and MOLLI were found for the basal anterior, mid anterolateral, apical anterior and apical lateral segments.

In terms of variance across subjects, the MOLLI estimates suggest large variance overall, with the largest difference to FLASH observed in the basal anterolateral segment (38.4 ms). These differences are likely a product of MOLLI's sharper $T_2$ maps, where pixels with lower $T_2$ values can be immediate neighbours of pixels with larger values, whereas in FLASH these transitions are smoother. This makes the MOLLI method more sensitive to segmentation errors, which may help explain the larger standard deviation observed. Furthermore, MOLLI's $T_2$ estimation is more sensitive to noise, which affects pixels differently.

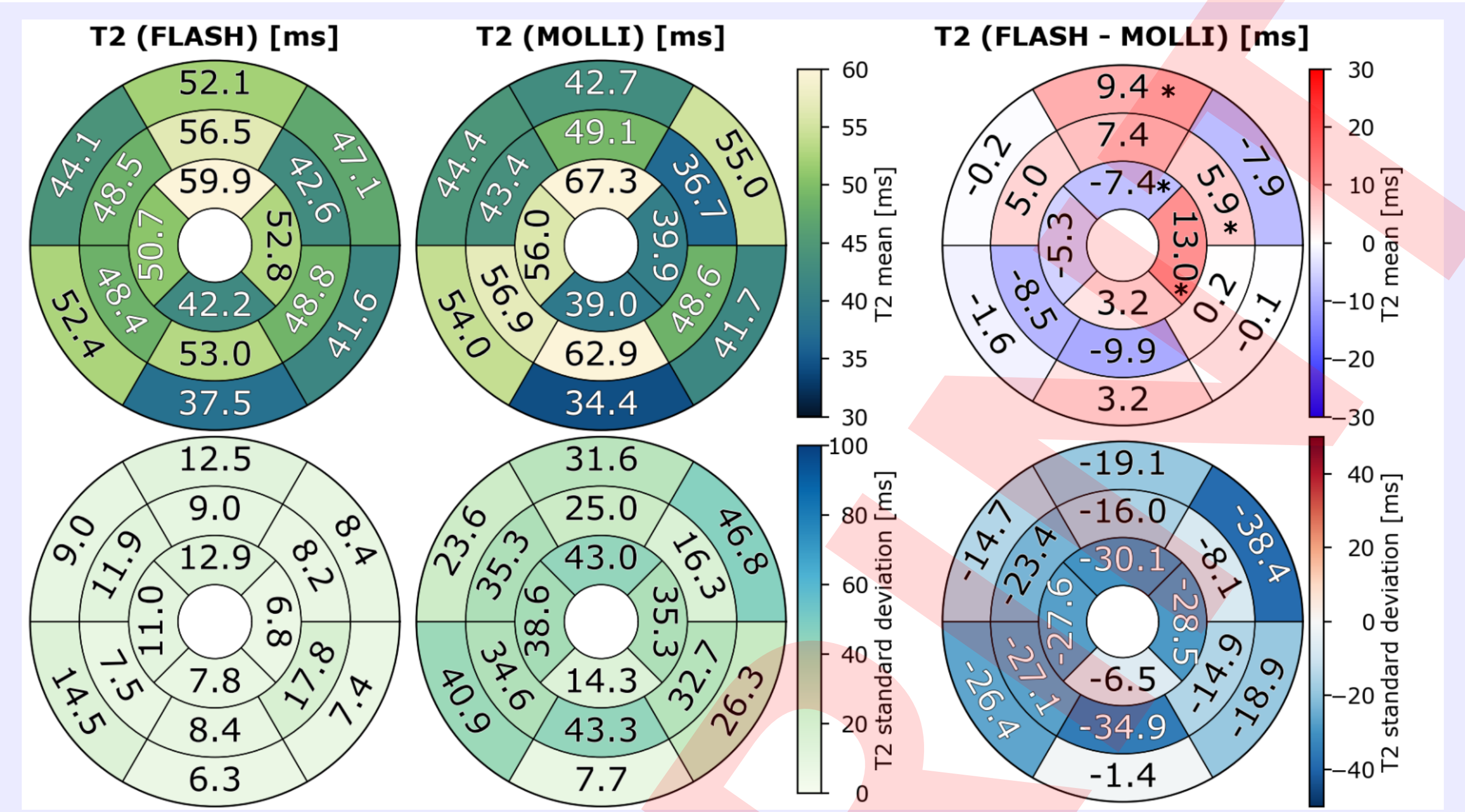

**Figure 7.** Segmental distribution of $T_2$ values estimated from MOLLI and FLASH, according to the AHA 16-segment model. **(top)** Mean $T_2$ values in each segment, averaged across all subjects. **(bottom)** Standard deviation of the $T_2$ values in each segment, averaged across all subjects. Significant errors are highlighted with $\star$ (p-value $< 0.05$).

## 4. Conclusions

The MOLLI acquisition sequence was developed for $T_1$ mapping. In this work, we evaluated MOLLI's sensitivity to $T_2$ and its ability to provide $T_2$ map estimates from $T_1$-weighted images, resorting to the EPG formulation, despite MOLLI not being specifically developed for $T_2$ mapping. Particularly, MOLLI's $T_2$ estimates are more accurate the lower the $T_2$ value. In the myocardium region, provided SNR is sufficiently high, MOLLI was able to accurately estimate $T_2$ values from the simulations, whereas in the blood regions or when there is a large amount of noise in the $T_1$-weighted images this ceased to be true. Nevertheless, this could allow clinicians to simultaneously assess myocardial $T_1$ and $T_2$ from a single acquisition sequence and use the accrued time for other analysis. It remains to be tested whether these results remain consistent across different MOLLI acquisition schemes, and whether they may prove useful in case of pathology.

## Abbreviations

| | |
|---|---|
| **AHA** | American Heart Association |
| **bSSFP** | Balanced Steady State Free Precession |
| **EPG** | Extended Phase Graph |
| **FLASH** | Fast Low Angle Shot |
| **FOV** | Field of View |
| **GRAPPA** | GeneRalized Autocalibrating Partial Parallel Acquisition |
| **HR** | Heart Rate |
| **MOLLI** | Modified Look-Locker Inversion Recovery |
| **MR** | Magnetic Resonance |
| **MRI** | Magnetic Resonance Imaging |
| **ROI** | Region of Interest |
| **SA** | Short Axis |
| **SNR** | Signal to Noise Ratio |